\documentclass[
10pt,
aps,
prb,
letterpaper,
twocolumn,
superscriptaddress,
preprintnumbers,%
floatfix,%
twoside]{revtex4-1}

\usepackage[usenames]{color}
\usepackage{times}
\usepackage{dcolumn}
\usepackage{bm}
\usepackage{ifthen}
\usepackage{amssymb}
\usepackage{amsfonts}
\usepackage{amsmath}
\usepackage{float}

\usepackage{graphicx}
\usepackage{float}
\usepackage{dcolumn}
\usepackage{bm}
\usepackage[colorlinks=true,linkcolor=blue, citecolor=blue]{hyperref}
\usepackage[mathlines]{lineno}
\usepackage{xcolor}
\usepackage[T1]{fontenc}
\usepackage{lettrine}
\usepackage[export]{adjustbox}
\usepackage{setspace}

\makeatletter
\renewenvironment{widetext@grid}{%
 \par\ignorespaces
 \setbox\widetext@top\vbox{%
  \vskip15\p@
  \hb@xt@\hsize{%
  \leaders\hrule\hfil
  \vrule\@height6\p@
  }%
  \vskip6\p@
 }%
 \setbox\widetext@bot\hb@xt@\hsize{%
  \vrule\@depth6\p@
  \leaders\hrule\hfil
 }%
 \onecolumngrid
}{%
 \par
 \dimen@\ht\widetext@bot\advance\dimen@\dp\widetext@bot
 \cleaders\box\widetext@bot\vskip\dimen@
 \twocolumngrid\global\@ignoretrue
 \@endpetrue
}%

\makeatother

\def\figurename{\textbf{\textsf{Fig.}}}
\renewcommand \thefigure {\textbf{\textsf{\arabic{figure}}}}

\newcommand\myCaption[2]{\begin{flushleft}\fontsize{7}{10}\selectfont\refstepcounter{figure}%
\noindent\textsf{\figurename}~\fontsize{7}{10}\selectfont\thefigure~\fontsize{7}{10}\selectfont{\textbf{\textsf{#1}}}
\textsf{#2}\end{flushleft}\small\selectfont}

\begin{document}

\title{\huge{\textrm{Hierarchies of Hofstadter butterflies in 2D covalent-organic frameworks}}}

\author{\large{David Bodesheim}}
\affiliation{Institute for Materials Science and Max Bergmann Centre of Biomaterials, Dresden University of Technology, 01069, Dresden Germany}

\author{\large{Robert Biele}}
\affiliation{Institute for Materials Science and Max Bergmann Centre of Biomaterials, Dresden University of Technology, 01069, Dresden Germany}

\author{\large{Gianaurelio Cuniberti\thanks{Corresponding author}}}
\email{gianaurelio.cuniberti@tu-dresden.de}
\affiliation{Institute for Materials Science and Max Bergmann Centre of Biomaterials, Dresden University of Technology, 01069, Dresden Germany}
\affiliation{Dresden Center for Computational Materials Science (DCMS), Dresden University of Technology, 01069, Dresden Germany}

\begin{abstract}
The Hofstadter butterfly is one of the first and most fascinating examples of the fractal and self-similar quantum nature of free electrons in a lattice pierced by a perpendicular magnetic field. However, the direct experimental verification of this effect on single-layer materials is still missing as very strong and inaccessible magnetic fields are necessary. For this reason, its indirect experimental verification has only been realized in artificial periodic 2D systems, like moiré lattices. The only recently synthesized 2D covalent-organic frameworks might circumvent this limitation: Due to their large pore structures, magnetic fields needed to detect most features of the Hofstadter butterfly are indeed accessible with today technology. This work opens the door to make this exotic and theoretical issue from the 70s measurable and might solve the quest for the experimental verification of the Hofstadter butterfly in single-layer materials. Moreover, the intrinsic hierarchy of different pore sizes in 2D covalent-organic framework adds additional complexity and beauty to the original butterflies and leads to a direct accessible playground for new physical observations.
\end{abstract}
\maketitle

\section*{INTRODUCTION}
The quantum behavior of an electron moving in a two-dimensional lattice exposed to a magnetic field is a fundamental problem of solid-state physics that has been studied deeply in the 70s. By plotting for the first time the allowed energies of the electrons and varying the magnetic field, Douglas Hofstadter found the stunning form of a graph that consists of duplicates of itself, embedded infinitely deeply, and that encodes nature in such a complex but equally self-similar manner. \cite{Hofstadter_1976} This so-called Hofstadter butterfly (HB), which is shown in Fig.~\ref{fig:Intro_fig}~a and b, emerges from the commensurability of the magnetic and the lattice lengths when constraining a free electron onto a lattice. This effect has fascinated theoretical physicists and mathematicians over many decades \cite{Rammal_1985, Gumbs_1997, Koshino_2001, Osadchy_2001, Xiao_2003, Nemec_2006, Nemec_2007, Ylmaz_2017,Ferrari_2015} and connects the mathematical concept of fractals with the world of physics by encoding the most exotic behaviors of quantum mechanics and topology, namely the quantum Hall effect and conductance quantization in 2D materials.\cite{vonKlitzing_1986,Kumar_2011,Li_2016,Kjaergaard_2016} However, the minimal magnetic field strength, $B_\mathrm{p}$, for resolving HBs depends inversely on the plaquette area $A_\mathrm{p}$ of a lattice closed loop, such as a hexagon in a honeycomb lattice,
\begin{equation}
\label{eq:Bperiod}
B_\mathrm{p} = \frac{\Phi_\mathrm{0}}{A_\mathrm{p}}\:.
\end{equation}

where, $\Phi_\mathrm{0} = \frac{h}{2e}$ is the magnetic flux quantum. The HB repeats itself infinitely with a period of $B_\mathrm{p}$. For graphene, for instance, $B_\mathrm{p}$ is around 78~kT, making it inaccessible for experimental validation. Hence, the HB has served since its discovery as a physical playground of mainly theoretical interest but with no direct observation. However, in the last three decades experimental verification of the HB via artificial lattices with very large unit cells, such as moiré lattices or 2D electron gas lattices, has turned this once-theoretical issue into an active field in experimental solid-state physics nowadays.\cite{Kuhl_1998,Albrecht_2001, Zaric_2004, Geisler_2004,Ponomarenko_2013,Dean_2013,Hunt_2013,Yu_2014,Yang_2016,Ni_2019,Huber_2020,Lu_2021} It was only the escamotage of artificial lattices, like moiré patterns, resulting in large plaquette areas which enabled the first indirect experimental demonstration of HB through magnetotransport measurements.
Two-dimensional covalent-organic frameworks (2D COFs), however, have not yet been at the center of such investigations and we can show that they indeed could be the first single-layered material to directly observe HBs. COFs are crystalline covalently bound organic polymers that have first been successfully synthesized in 2005.\cite{Yaghi_2005} The crystalline polymers are constructed from organic building block molecules attached to each other in a regular fashion which creates a periodic and often porous structure as depicted in Fig.~\ref{fig:Intro_fig} c. Due to the many available types of precursor molecules, various different structures with tunable properties are possible. As a result, COFs have gained ever since a lot of attention in the polymer and solid state community.
\begin{widetext}
\begin{minipage}{\linewidth}
\begin{figure}[H]
\centering
\includegraphics[width=\linewidth]{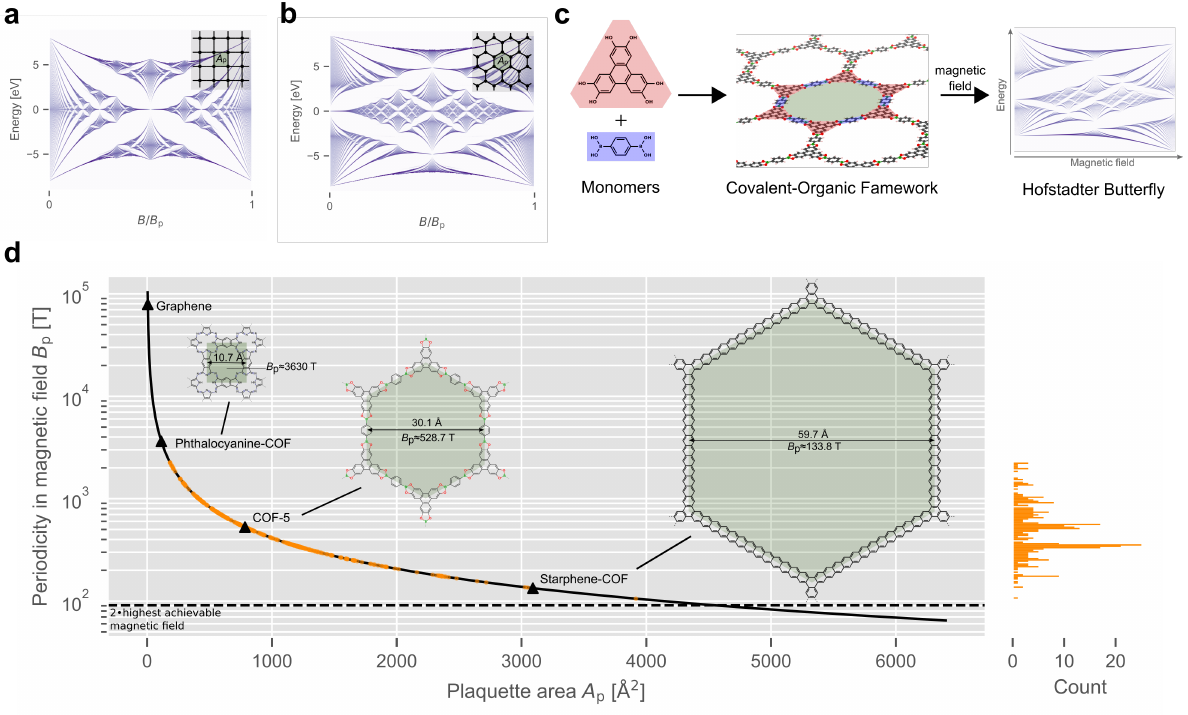}
\myCaption{From covalent-organic frameworks to a measurable Hofstadter butterflies.\label{fig:Intro_fig}}{\textbf{a} (and \textbf{b}) Application of a magnetic field to a square (honeycomb) lattice with a plaquette area $A_\mathrm{p}$, marked as light green, leads to the fractal HB pattern with a period of $B_\mathrm{p}$. \textbf{c} From left to right: the lattice formation of COF-5 from its tritopic (red) and linear (blue) molecular building blocks. Applying a perpendicular magnetic field to a COF leads to a HB resolved at small magnetic fields $B_\mathrm{p}$ \textbf{d} Relation between the magnetic field to resolve a full HB $B_\mathrm{p}$ and the area of a plaquette according to Eq. (\ref{eq:Bperiod}). The dashed black line indicates twice the highest continuous magnetic field achievable to this date.\cite{Hahn_2019} The orange points represent experimentally known COFs with square, tetragonal and honeycomb lattices collected from the CURATED-COFs database (for details on the curation, see SI).\cite{CURATED_2019} The black triangles indicate graphene and the three COFs investigated in this study whose structures are shown.
} 
\end{figure}
\end{minipage}
\end{widetext}

Many COFs are layered materials, similar to the graphene sheets in graphite, but experimental methods have been refined to create few-layer and indeed monolayer 2D COFs.\cite{Dienstmaier_2011,Dienstmaier_2012,Colson_2013,Zhong_2019,Liu_2019,Li_2020, Li2_2020, OrtegaGuerrero_2021} These are based on simpler lattice types, like honeycomb or hexagonal lattices, with usually very large unit cells. This would lead to a lower magnetic field required for the experimental validation of the HB. Additionally, it has been shown how in these materials the lattice topology itself mediates electronic and mechanical properties.\cite{Springer_2020, Raptakis_2021} Moreover, recent investigations of topological effects in 2D COFs showed their potential as topological materials. \cite{Dong_2016,Jiang_2021,Pham_2021,Cui_2020, Li_Lei_2020, Jiang_2020} All of this makes 2D COFs interesting materials for the exploration of fractality, quantum effects and finally finding a HB at experimentally viable magnetic fields.

In Fig.~\ref{fig:Intro_fig}~d, the dependency of $B_\mathrm{p}$ on the plaquette area $A_\mathrm{p}$ is depicted, including known 2D COFs with honeycomb, tetragonal or square lattice structure. This shows that several COFs already exist where the HB can be measured to a great extent with experimentally achievable magnetic fields. We chose $91~\mathrm{T}$ as comparison, which is twice the highest achievable continuous magnetic field ($2\cdot45.5~\mathrm{T} = 91~\mathrm{T}$ \cite{Hahn_2019}), as the HB is symmetric in the magnetic field and consequently only half of its spectrum needs to be measured. Additionally, a greater complexity and a richer phenomenon of the HB is expected due to COFs rich pore hierarchy. The HB for three different exemplary 2D COFs (shown in Fig. \ref{fig:Intro_fig}~d) are calculated in the following: complex hierarchical patterns and fractal HBs can be observed at measurable magnetic fields.\\

\begin{widetext}
\begin{minipage}{\linewidth}
\begin{figure}[H]
\centering
\includegraphics[width=\linewidth]{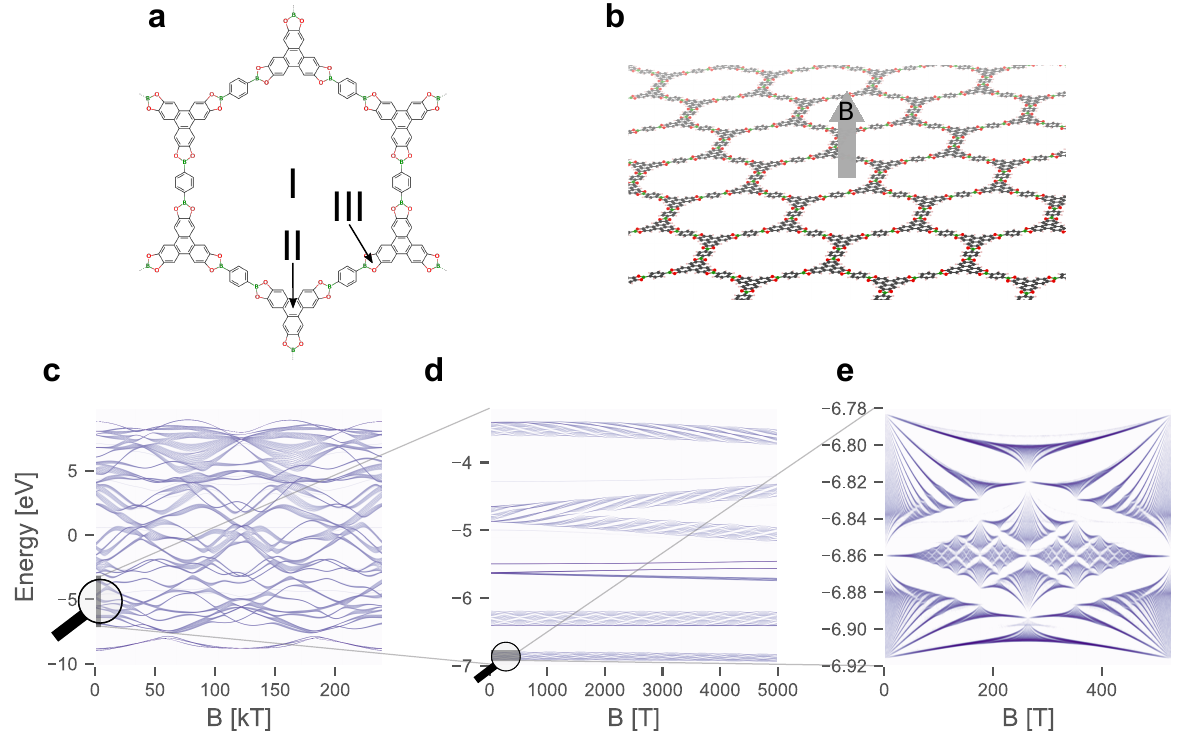}
\myCaption{Hierarchical periodic spectra for COF-5.\label{fig:COF-5_HSBFs}}{ \textbf{a} Structure of COF-5 including annotations for the largest to the smallest plaquettes (\textbf{I}-\textbf{III}). \textbf{b} Extended 2D structure of the honeycomb lattice of COF-5 with indicated applied magnetic field. \textbf{c} Butterfly spectrum from 0 to 240~kT. \textbf{d} Butterfly spectrum from 0 to 5000~T. \textbf{e} Honeycomb HB from 0 to 530~T. The intensity of the DOS in the butterfly spectra is depicted in a logarithmic scale although the colorbar was omitted for clarity.}

\end{figure}
\end{minipage}
\end{widetext}

\section*{RESULTS}
\subsection*{Embedded Hofstadter butterflies}
Fig.~\ref{fig:COF-5_HSBFs}~shows the well-known COF-5 with its three different plaquette types marked as \textbf{I} to \textbf{III}. COF-5 is the first synthesized and one of the most studied COF which has, just like graphene, a honeycomb lattice as shown in Fig.~\ref{fig:COF-5_HSBFs}~b.\cite{Yaghi_2005}
As the HB repeats itself with a period of $B_\mathrm{p}$, we search for a reoccurring pattern in the calculated spectrum. In Fig. \ref{fig:COF-5_HSBFs}~c, one can see such a pattern with a period of approximately 240~kT. This spectrum, however, does not resemble the HB of an equivalent honeycomb lattice (Fig. \ref{fig:Intro_fig}~b) and its period is very large. A zoom into the spectrum, as shown in Fig. \ref{fig:COF-5_HSBFs}~d, indicate more complex patterns and a further zoom in Fig. \ref{fig:COF-5_HSBFs}~e reveals astonishingly a HB that is equivalent of the one of graphene, but only at 528.7~T. This corresponds to the $B_\mathrm{p}$ of the lattice plaquette \textbf{I} according to equation~(\ref{eq:Bperiod}). To avoid confusion in the following discussion, we will call the patterns that resemble the HB of simple lattices specifically "HB" and the rest of the spectrum generally "butterfly spectrum".

How can we understand the formation of different patterns on the various scales of magnetic fields? As seen in Fig.~\ref{fig:COF-5_HSBFs}~c, the butterfly spectrum contains many different "bands" that change with the magnetic field. We will call those "ribbons" to give a clear distinction to bands in band structure plots. This complexity arises due to the numerous electronic states that COFs have compared to simple lattices. 

A zoom closer to some of the ribbons in Fig.~\ref{fig:COF-5_HSBFs}~d, shows that each ribbon itself contains many smaller complex features that sometimes overlap and interfere with each other and create new patterns. 
\begin{widetext}
\begin{minipage}{\linewidth}
\begin{figure}[H]
\centering
\includegraphics[width=\linewidth]{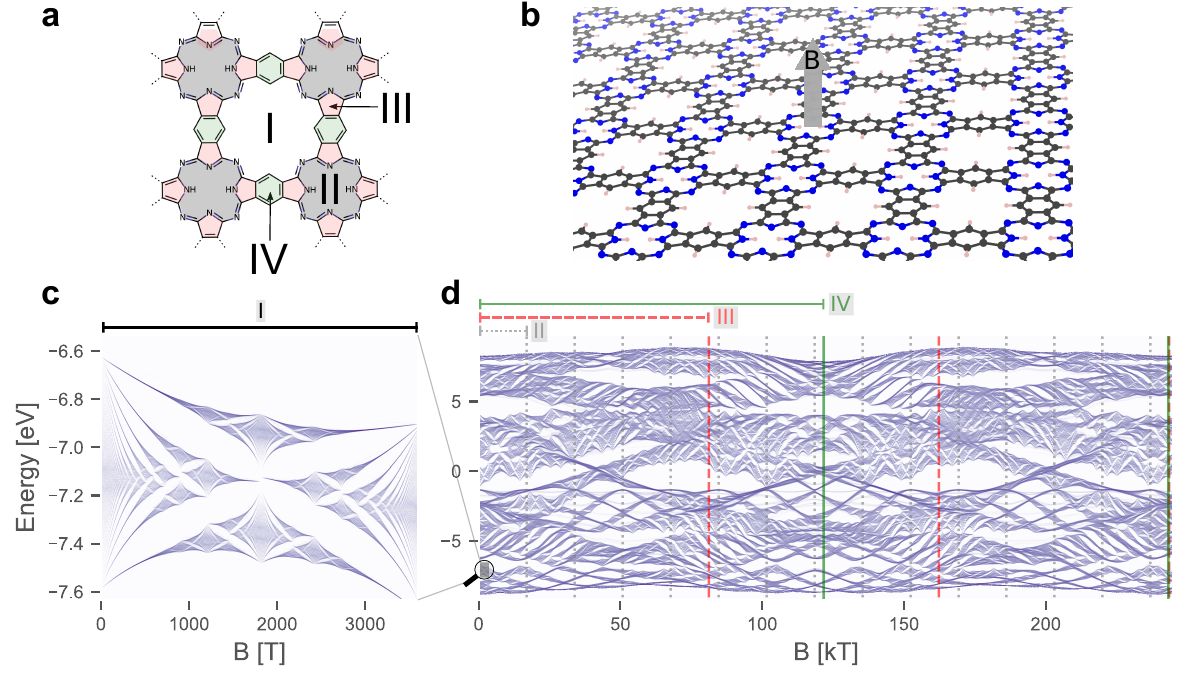}
\myCaption{Different periodicities in the butterfly spectrum shown in the example of Phthalocyanine-COF.\label{fig:Phthalocyanine_HSBFs}}{\textbf{a} Structure of the Phthalocyanine-COF including annotations for the largest to the smallest plaquettes (\textbf{I}-\textbf{IV}). \textbf{b} Extended 2D structure of Phthalocyanine-COF including an indicated applied magnetic field. \textbf{c} HB from 0 to 3600~T. \textbf{d} Butterfly spectrum from 0 to 245~kT. The solid green lines indicate the periodicity generated by plaquette IV (121.6~kT), the dashed red line by plaquette III (81.18~kT) and the dotted gray line by plaquette \textbf{II} (16.92~kT). The intensity of the DOS in the butterfly spectra is depicted in a logarithmic scale although the colorbar was omitted for clarity.}
\end{figure}
\end{minipage}
\end{widetext}

A comparison with the band structure in SI~Fig.~2 reveals that some ribbons are based on a band-structure equivalent of a simple honeycomb lattice, while others are more similar to the one of a kagome lattice. It is known that COF-5 has kagome-characteristic bands.\cite{Kuc_2020} The resulting butterfly spectrum of these kagome-like bands is similarly distorted as the one of a simple kagome-lattice HB.\cite{Xiao_2003} Furthermore, a comparison of the band-structure and HBs of all studied COFs in this manuscript is shown in SI~Fig.~5.
The final zoom in Fig.~\ref{fig:COF-5_HSBFs}~e into the isolated ribbon at around -6.9~eV shows the honeycomb HB. This shows that the lattice type itself dictates the electronic structure and topological effects. A key difference to the HB in a simple honeycomb lattice is that HB in COF-5 is strictly speaking not repeating itself, but is distorted for higher magnetic fields due to the superposition of different periodicities.\\

\subsection*{Hierarchical patterns in 2D COFs}
But why do we see a periodic pattern at approximately 240~kT in Fig.~\ref{fig:COF-5_HSBFs}~c for COF-5? This value does not coincide with any of the $B_\mathrm{p}$ values of the three plaquette types \textbf{I} to \textbf{III} in COF-5 labelled in Fig.~\ref{fig:COF-5_HSBFs}~a: the honeycomb lattice plaquette \textbf{I}, the benzene ring \textbf{II} and the boronic ester based five-membered ring \textbf{III}, with periodicities $B_\mathrm{p}$ of 528.7~T, 81.06~kT and 121.9~kT respectively. As the latter two periods have approximately a common denominator, a new periodic pattern with a bigger period at approximately 240~kT ($\approx 2\cdot 121.9~\mathrm{kT} \approx 3\cdot 81~\mathrm{kT}$) is visible.

To better understand the influence of multiple plaquette types on the butterfly spectrum, we next investigate a COF with four different plaquettes. This COF is based on Phthalocyanine as shown in Fig. \ref{fig:Phthalocyanine_HSBFs}~a and has a square lattice. It has been shown in previous studies that this COF has interesting topological properties, although usually investigated including a metal-center.\cite{Jiang_2020, Li_Lei_2020, Pham_2021} In Fig. \ref{fig:Phthalocyanine_HSBFs}~a, we depict the four different kinds of plaquettes marked as \textbf{I} to \textbf{IV}: square lattice plaquette that arises from the fused Phthalocyanines units \textbf{I}, the pore inside of a Phthalocyanine unit \textbf{II}, the benzene ring \textbf{III} and the pyrole five-membered ring \textbf{IV}. In Fig.~\ref{fig:Phthalocyanine_HSBFs}~d, we can see that similar to COF-5, there is an overlap of the periodicities of \textbf{III} (81.18~kT) and \textbf{IV} (121.6~kT) that lead to a new combined pattern with a period of approximately 240~kT.

\begin{widetext}
\begin{minipage}{\linewidth}
\begin{figure}[H]
\centering
\includegraphics[width=\linewidth]{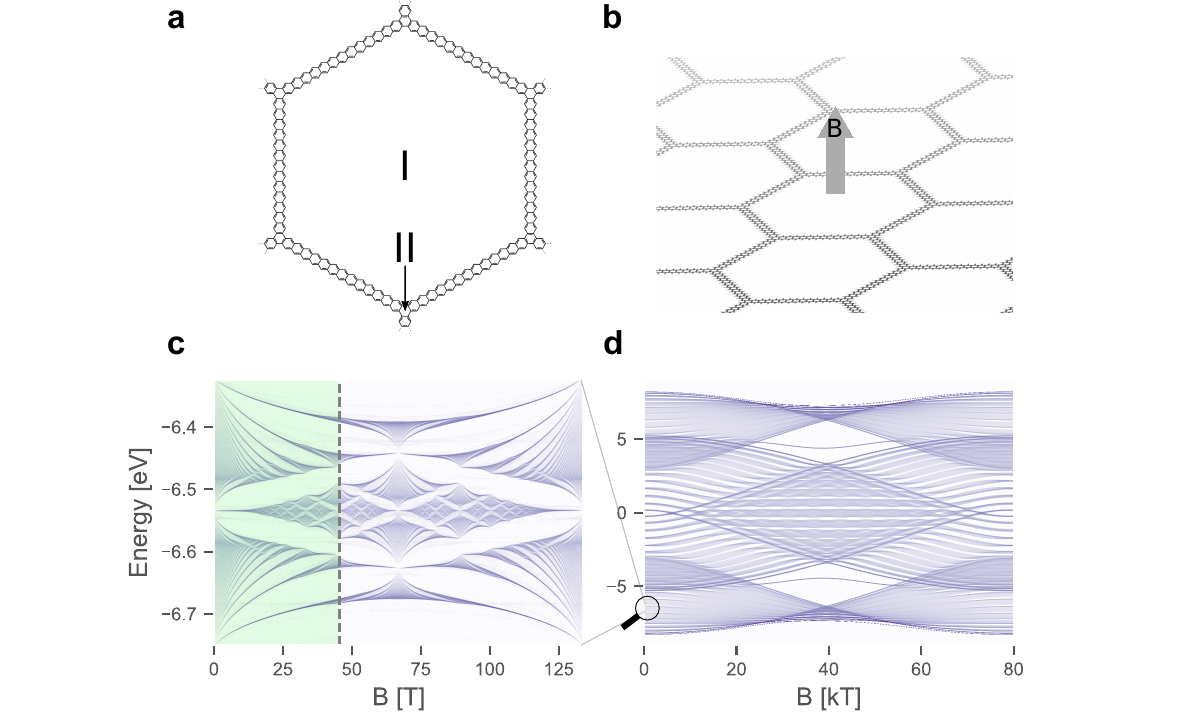}
\myCaption{Towards a measurable HB in 2D COFs with the Starphene-COF.\label{fig:Starphene_HSBFs}}{ \textbf{a} Structure of the Starphene-COF. \textbf{b} Extended 2D structure of the honeycomb lattice of the Starphene-COF including an indicated applied magnetic field. \textbf{c} HB from 0 to 134~T. The green overlay indicates the range in which the spectrum would be measurable according to the highest achievable magnetic field of 45.5~T.\cite{Hahn_2019} \textbf{d} Butterfly spectrum from 0 to 80~kT. The intensity of the DOS is depicted in a logarithmic scale although the colorbar was omitted for clarity.}
\end{figure}
\end{minipage}
\end{widetext}

The spectrum shows that there are patterns with different periods which are marked in the figure. For better visualization of the repeating patterns, a wider magnetic field range of the butterfly spectrum plotted in SI~Fig.~3. When zooming into one of the ribbons in Fig. \ref{fig:Phthalocyanine_HSBFs}~d, we see as expected a HB in Fig. \ref{fig:Phthalocyanine_HSBFs}~c that resembles the one of a square lattice (compare to Fig. \ref{fig:Intro_fig}~a, which further shows how lattice types guides topological effects like the HB. But since the period of \textbf{II} is only around five times larger than \textbf{I}, the oscillation of \textbf{II} distrorts the HB in Fig.~\ref{fig:Phthalocyanine_HSBFs}~c. From this we can see that the closer the scales of periodicities of different plaquettes are, the more they influence each other and the more distorted the HB is.

We would like to point out that the fractality and the actual HB is only created by the lattice plaquette, i.e. the biggest plaquette. The other plaquettes are only responsible for additional periodicity, but not for fractality, as shown in SI~Fig.~4.\\

\subsection*{Towards a measurable HB in 2D COFs}
The COFs investigated so far do not have large enough unit cells to create a periodicity $B_\mathrm{p}$ that would be measurable in today's experiments. Hence, a COF with larger pores is required. One COF (COF-122) exists already that is close to the required unit cell size.\cite{Zhao_2020} However, as parts of this COF have many rotational degrees of freedom and is hence not flat, it is in our opinion not suitable for a flat monolayer material. For this reason, we propose a COF that is based on the recently reported class of Starphene molecules which are triangular molecules of fused benzene rings.\cite{Holec_2021} When connected, a highly porous COF with a honeycomb lattice would result which we call Starphene-COF, as shown in Fig.~\ref{fig:Starphene_HSBFs}~a and b. This theoretical COF also is a fully conjugated system which should lead to a good conductivity. In the Starphene-COF, there are only two plaquette types: the honeycomb lattice plaquette \textbf{I} and the benzene-rings \textbf{II}. This leads to a period of 133.8~T for the lattice plaquette \textbf{I} and a period of approximately 78.68~kT for the benzene ring (\textbf{II}). The full butterfly spectrum up to 80~kT is depicted in Fig.~\ref{fig:Starphene_HSBFs}~d. In Fig. \ref{fig:Starphene_HSBFs}~c, we zoom in to the magnetic field range of the smaller period and see that a HB arises that looks similar to the one of a simple honeycomb lattice in Fig. \ref{fig:Intro_fig}~b. Since $B_\mathrm{p}$ is only 133.8~T, the HB of the Starphene-COF would be accessible to a great extent by available magnetic fields, possibly solving the quest for direct experimental verification of the HB in a single layer 2D material in the future.

The butterfly spectrum in Fig.~\ref{fig:Starphene_HSBFs}~d also has some interesting properties similar to a related class of materials, graphene antidot lattices (GALs) which is graphene with periodic holes. In GALs, a typical graphene HB pattern still emerges, even though it becomes less distinct the bigger the holes in the lattice becomes. The Starphene-COF can be seen as an edge case of such a GAL, in the sense of a porous graphene which has the biggest possible hole for its repeating unit. Another feature that is known from GALs is that the band gap is being quenched with increasing magnetic field, which can be seen for the Starphene-COF at around 6~kT at 0~eV in Fig.~\ref{fig:Starphene_HSBFs}~d.\cite{Pedersen_2013}

During the preparation of the manuscript, a novel COF with a very large pore was reported.\cite{Riao_2021} We show in SI~Fig.~1 the corresponding HB that is resolved at 122.6~T, which is even lower than our proposed Starphene-COF. This demonstrates how the rapid advances in synthesis can provide systems to experimentally measure the HB in the near future. \\

\section*{DISCUSSION}
We have shown that 2D COFs are promising materials for the direct experimental verification of the Hofstadter butterfly in pure 2D materials and that there are additional hierarchical patterns that are not present in HBs of simple lattices. This work might turn this merely-theoretical issue from the 70s into an active experimental research activity nowadays and trigger the race for the first direct measurement.
We found equivalents of simple lattice HB inside the complex and beautiful butterfly spectra of 2D COFs. Moreover, hierarchical structures and periodicities can be found within these spectra, due to the various pore types inside a single COF-structure, opening a new dimension for novel physics. Furthermore, we showed that for 2D COFs with large unit cells the magnetic field to measure a HB is small enough to be experimentally viable. 2D COFs can provide the ideal playground for the investigation of such fundamental quantum effects and the recent advances in single-layer synthesis of 2D COFs will lead in the foreseeable future to the direct observation of the Hofstadter butterfly in a single-layer 2D material. \\

\section*{METHODS}
{\scriptsize
\subsection*{Calculation of the HB}
{A tight-binding hamiltonian was constructed from the \textit{Slater-Koster} parametrization which is used in Density Functional based Tight Binding. The parameters include tabulated distance-dependent orbital-orbital interactions and respective on-site energies.\cite{Slater_1954} Here, the \textit{matsci} parametrization with on-site energies for the angular momentum p for the given atom and $\mathrm{pp^{\pi}}$ hopping parameters.\cite{Manzano_2012}} The hopping was assigned according to the respective distance between the atoms. To simplify the model and the energy-spectrum, only nearest neighbor interactions were considered. With the help of the \textit{Atomic Simulation Environment} (\textit{ASE}) package,\cite{ASE_2017} the geometry and hamiltonian were imported into the \textit{pybinding} code with which all further calculations were performed.\cite{pybinding} For the geometry, only a strict 2D system was used, meaning the z-coordinates were discarded. The influence of a perpendicular magnetic field on the 2D structure was approximated with the Peierls-substitution:\cite{Peierls_1933}

\begin{equation}
  t_{nm} \to t_{nm} \mathrm{e}^{i\frac{2\pi}{\Phi_0}\int_{n}^{m} \vec{A}_{nm} \mathrm{d}\vec{l}}
\end{equation}

with $t_{nm}$ as the hopping-parameter or off-diagonal elements for nearest neighbor sites $n$ and $m$, $\Phi_0$ as the magnetic flux quantum, $\vec{A}_{nm}$ as the magnetic vector potential with a chosen gauge of $\vec{A}(x,y,z)=(B_y,0,0)$. The electronic density of states (DOS) for the structures with an applied perpendicular magnetic field was calculated with the in \textit{pybinding} integrated Kernel Polynomial Method (KPM). In the KPM, the DOS is calculated by reconstruction from the KPM moments which are evaluated in a stochastic fashion. \cite{KPM_method} The DOS was evaluated for a 50x50 supercell and 3 random vectors for the stochastic calculation of KPM moments. Sweeping through a magnetic field range and plotting the DOS for each magnetic field, results in the shown butterfly spectra. The energy scale was centered with respect to the the minimum and maximum energy value of the butterfly spectrum. This convention for the energy scale was chosen instead Fermi level centering because the TB model only includes $\mathrm{pp^{\pi}}$ orbitals and their occupation for more complex systems (like the Phthalocyanine-COF or COF-5) cannot be derived directly from the TB model.
\\

\subsection*{Optimization of COF-Structures}

The 2D COF-structures where obtained from optimization with Density Functional based Tight Binding calculations with the code \textit{DFTB+}.\cite{DFTB_2020} The optimization was carried out at the $\Gamma$ point with a maximum force component of 1e-06 of the conjugate gradient algorithm. A tolerance of 1e-08 for the self consistent cycles was chosen. As \textit{Slater-Koster} parameters, the \textit{matsci} parameter set was used.\cite{Manzano_2012}\\
}

\noindent \textsf{\textbf{Data availability}}\\
The data necessary to produce the butterfly spectrum is uploaded to \url{https://github.com/DBodesheim/HB_pybinding} upon publication.\\\\
\noindent \textsf{\textbf{Code availability}}\\
The code necessary to produce the butterfly spectrum is uploaded to \url{https://github.com/DBodesheim/HB_pybinding} upon publication.\\\\

\bibliographystyle{unsrt}
\bibliography{references}

\begin{thebibliography}{1}

\bibitem{Riao_2021}
Alberto Ria{\~{n}}o, Karol Struty{\'{n}}ski, Meng Liu, Craig~T. Stoppiello,
  Bel{\'{e}}n Lerma-Berlanga, Akinori Saeki, Carlos Mart{\'{\i}}-Gastaldo,
  Andrei~N. Khlobystov, Giovanni Valenti, Francesco Paolucci, Manuel
  Melle-Franco, and Aurelio Mateo-Alonso.
\newblock An expanded 2d fused aromatic network with 90-ring hexagons.
\newblock {\em Angewandte Chemie International Edition}, December 2021.

\bibitem{Hahn_2019}
Seungyong Hahn, Kwanglok Kim, Kwangmin Kim, Xinbo Hu, Thomas Painter, Iain
  Dixon, Seokho Kim, Kabindra~R. Bhattarai, So~Noguchi, Jan Jaroszynski, and
  David~C. Larbalestier.
\newblock 45.5-tesla direct-current magnetic field generated with a
  high-temperature superconducting magnet.
\newblock {\em Nature}, 570(7762):496--499, June 2019.

\bibitem{CURATED_2019}
Daniele Ongari, Aliaksandr~V. Yakutovich, Leopold Talirz, and Berend Smit.
\newblock Building a consistent and reproducible database for adsorption
  evaluation in covalent{\textendash}organic frameworks.
\newblock {\em {ACS} Central Science}, 5(10):1663--1675, September 2019.

\end{thebibliography}


\begin{thebibliography}{10}

\bibitem{Hofstadter_1976}
Douglas~R. Hofstadter.
\newblock Energy levels and wave functions of bloch electrons in rational and
  irrational magnetic fields.
\newblock {\em Phys. Rev. B}, 14:2239--2249, Sep 1976.

\bibitem{Rammal_1985}
R.~Rammal.
\newblock Landau level spectrum of bloch electrons in a honeycomb lattice.
\newblock {\em Journal de Physique}, 46(8):1345--1354, 1985.

\bibitem{Gumbs_1997}
Godfrey Gumbs and Paula Fekete.
\newblock Hofstadter butterfly for the hexagonal lattice.
\newblock {\em Physical Review B}, 56(7):3787--3791, August 1997.

\bibitem{Koshino_2001}
M.~Koshino, H.~Aoki, K.~Kuroki, S.~Kagoshima, and T.~Osada.
\newblock Hofstadter butterfly and integer quantum hall effect in three
  dimensions.
\newblock {\em Physical Review Letters}, 86(6):1062--1065, February 2001.

\bibitem{Osadchy_2001}
D.~Osadchy and J.~E. Avron.
\newblock Hofstadter butterfly as quantum phase diagram.
\newblock {\em Journal of Mathematical Physics}, 42(12):5665--5671, December
  2001.

\bibitem{Xiao_2003}
Yi~Xiao, Vincent Pelletier, Paul~M. Chaikin, and David~A. Huse.
\newblock Landau levels in the case of two degenerate coupled
  bands:{\hspace{0.6em}}kagom{\'{e}} lattice tight-binding spectrum.
\newblock {\em Physical Review B}, 67(10), March 2003.

\bibitem{Nemec_2006}
Norbert Nemec and Gianaurelio Cuniberti.
\newblock Hofstadter butterflies of carbon nanotubes: Pseudofractality of the
  magnetoelectronic spectrum.
\newblock {\em Phys. Rev. B}, 74:165411, Oct 2006.

\bibitem{Nemec_2007}
Norbert Nemec and Gianaurelio Cuniberti.
\newblock Hofstadter butterflies of bilayer graphene.
\newblock {\em Phys. Rev. B}, 75:201404, May 2007.

\bibitem{Ylmaz_2017}
F.~Y{\i}lmaz and M.~\"{O}. Oktel.
\newblock Hofstadter butterfly evolution in the space of two-dimensional
  bravais lattices.
\newblock {\em Physical Review A}, 95(6), June 2017.

\bibitem{Ferrari_2015}
Andrea~C. Ferrari, Francesco Bonaccorso, Vladimir Fal{\textquotesingle}ko,
  Konstantin~S. Novoselov, Stephan Roche, Peter B{\o}ggild, Stefano Borini,
  Frank H.~L. Koppens, Vincenzo Palermo, Nicola Pugno, Jos{\'{e}}~A. Garrido,
  Roman Sordan, Alberto Bianco, Laura Ballerini, Maurizio Prato, Elefterios
  Lidorikis, Jani Kivioja, Claudio Marinelli, Tapani Ryh\"{a}nen, Alberto
  Morpurgo, Jonathan~N. Coleman, Valeria Nicolosi, Luigi Colombo, Albert Fert,
  Mar Garcia-Hernandez, Adrian Bachtold, Gr{\'{e}}gory~F. Schneider, Francisco
  Guinea, Cees Dekker, Matteo Barbone, Zhipei Sun, Costas Galiotis,
  Alexander~N. Grigorenko, Gerasimos Konstantatos, Andras Kis, Mikhail
  Katsnelson, Lieven Vandersypen, Annick Loiseau, Vittorio Morandi, Daniel
  Neumaier, Emanuele Treossi, Vittorio Pellegrini, Marco Polini, Alessandro
  Tredicucci, Gareth~M. Williams, Byung~Hee Hong, Jong-Hyun Ahn, Jong~Min Kim,
  Herbert Zirath, Bart~J. van Wees, Herre van~der Zant, Luigi Occhipinti,
  Andrea~Di Matteo, Ian~A. Kinloch, Thomas Seyller, Etienne Quesnel, Xinliang
  Feng, Ken Teo, Nalin Rupesinghe, Pertti Hakonen, Simon R.~T. Neil, Quentin
  Tannock, Tomas L\"{o}fwander, and Jari Kinaret.
\newblock Science and technology roadmap for graphene, related two-dimensional
  crystals, and hybrid systems.
\newblock {\em Nanoscale}, 7(11):4598--4810, 2015.

\bibitem{vonKlitzing_1986}
Klaus von Klitzing.
\newblock The quantized hall effect.
\newblock {\em Reviews of Modern Physics}, 58(3):519--531, July 1986.

\bibitem{Kumar_2011}
A.~Kumar, W.~Escoffier, J.~M. Poumirol, C.~Faugeras, D.~P. Arovas, M.~M.
  Fogler, F.~Guinea, S.~Roche, M.~Goiran, and B.~Raquet.
\newblock Integer quantum hall effect in trilayer graphene.
\newblock {\em Physical Review Letters}, 107(12), September 2011.

\bibitem{Li_2016}
Likai Li, Fangyuan Yang, Guo~Jun Ye, Zuocheng Zhang, Zengwei Zhu, Wenkai Lou,
  Xiaoying Zhou, Liang Li, Kenji Watanabe, Takashi Taniguchi, Kai Chang, Yayu
  Wang, Xian~Hui Chen, and Yuanbo Zhang.
\newblock Quantum hall effect in black phosphorus two-dimensional electron
  system.
\newblock {\em Nature Nanotechnology}, 11(7):593--597, March 2016.

\bibitem{Kjaergaard_2016}
M.~Kjaergaard, F.~Nichele, H.~J. Suominen, M.~P. Nowak, M.~Wimmer, A.~R.
  Akhmerov, J.~A. Folk, K.~Flensberg, J.~Shabani, C.~J. Palmstr{\o}m, and C.~M.
  Marcus.
\newblock Quantized conductance doubling and hard gap in a two-dimensional
  semiconductor{\textendash}superconductor heterostructure.
\newblock {\em Nature Communications}, 7(1), September 2016.

\bibitem{Kuhl_1998}
U.~Kuhl and H.-J. St\"ockmann.
\newblock Microwave realization of the hofstadter butterfly.
\newblock {\em Phys. Rev. Lett.}, 80:3232--3235, Apr 1998.

\bibitem{Albrecht_2001}
C.~Albrecht, J.~H. Smet, K.~von Klitzing, D.~Weiss, V.~Umansky, and
  H.~Schweizer.
\newblock Evidence of hofstadter's fractal energy spectrum in the quantized
  hall conductance.
\newblock {\em Phys. Rev. Lett.}, 86:147--150, Jan 2001.

\bibitem{Zaric_2004}
Sasa Zaric, Gordana~N. Ostojic, Junichiro Kono, Jonah Shaver, Valerie~C. Moore,
  Michael~S. Strano, Robert~H. Hauge, Richard~E. Smalley, and Xing Wei.
\newblock Optical signatures of the aharonov-bohm phase in single-walled carbon
  nanotubes.
\newblock {\em Science}, 304(5674):1129--1131, May 2004.

\bibitem{Geisler_2004}
M.~C. Geisler, J.~H. Smet, V.~Umansky, K.~von Klitzing, B.~Naundorf,
  R.~Ketzmerick, and H.~Schweizer.
\newblock Detection of a landau band-coupling-induced rearrangement of the
  hofstadter butterfly.
\newblock {\em Phys. Rev. Lett.}, 92:256801, Jun 2004.

\bibitem{Ponomarenko_2013}
L.~A. Ponomarenko, R.~V. Gorbachev, G.~L. Yu, D.~C. Elias, R.~Jalil, A.~A.
  Patel, A.~Mishchenko, A.~S. Mayorov, C.~R. Woods, J.~R. Wallbank,
  M.~Mucha-Kruczynski, B.~A. Piot, M.~Potemski, I.~V. Grigorieva, K.~S.
  Novoselov, F.~Guinea, V.~I. Fal'ko, and A.~K. Geim.
\newblock Cloning of dirac fermions in graphene superlattices.
\newblock {\em Nature}, 497(7451):594--597, May 2013.

\bibitem{Dean_2013}
C.~R. Dean, L.~Wang, P.~Maher, C.~Forsythe, F.~Ghahari, Y.~Gao, J.~Katoch,
  M.~Ishigami, P.~Moon, M.~Koshino, T.~Taniguchi, K.~Watanabe, K.~L. Shepard,
  J.~Hone, and P.~Kim.
\newblock Hofstadter's butterfly and the fractal quantum hall effect in
  moir{\'{e}} superlattices.
\newblock {\em Nature}, 497(7451):598--602, May 2013.

\bibitem{Hunt_2013}
B.~Hunt, J.~D. Sanchez-Yamagishi, A.~F. Young, M.~Yankowitz, B.~J. LeRoy,
  K.~Watanabe, T.~Taniguchi, P.~Moon, M.~Koshino, P.~Jarillo-Herrero, and R.~C.
  Ashoori.
\newblock Massive dirac fermions and hofstadter butterfly in a van der waals
  heterostructure.
\newblock {\em Science}, 340(6139):1427--1430, May 2013.

\bibitem{Yu_2014}
G.~L. Yu, R.~V. Gorbachev, J.~S. Tu, A.~V. Kretinin, Y.~Cao, R.~Jalil,
  F.~Withers, L.~A. Ponomarenko, B.~A. Piot, M.~Potemski, D.~C. Elias, X.~Chen,
  K.~Watanabe, T.~Taniguchi, I.~V. Grigorieva, K.~S. Novoselov, V.~I. Fal'ko,
  A.~K. Geim, and A.~Mishchenko.
\newblock Hierarchy of hofstadter states and replica quantum hall
  ferromagnetism in graphene superlattices.
\newblock {\em Nature Physics}, 10(7):525--529, June 2014.

\bibitem{Yang_2016}
Wei Yang, Xiaobo Lu, Guorui Chen, Shuang Wu, Guibai Xie, Meng Cheng, Duoming
  Wang, Rong Yang, Dongxia Shi, Kenji Watanabe, Takashi Taniguchi, Christophe
  Voisin, Bernard Plaçais, Yuanbo Zhang, and Guangyu Zhang.
\newblock Hofstadter butterfly and many-body effects in epitaxial graphene
  superlattice.
\newblock {\em Nano Letters}, 16(4):2387--2392, 2016.
\newblock PMID: 26950258.

\bibitem{Ni_2019}
Xiang Ni, Kai Chen, Matthew Weiner, David~J. Apigo, Camelia Prodan, Andrea
  Al{\`{u}}, Emil Prodan, and Alexander~B. Khanikaev.
\newblock Observation of hofstadter butterfly and topological edge states in
  reconfigurable quasi-periodic acoustic crystals.
\newblock {\em Communications Physics}, 2(1), June 2019.

\bibitem{Huber_2020}
Robin Huber, Ming-Hao Liu, Szu-Chao Chen, Martin Drienovsky, Andreas Sandner,
  Kenji Watanabe, Takashi Taniguchi, Klaus Richter, Dieter Weiss, and Jonathan
  Eroms.
\newblock Gate-tunable two-dimensional superlattices in graphene.
\newblock {\em Nano Letters}, 20(11):8046--8052, 2020.
\newblock PMID: 33054236.

\bibitem{Lu_2021}
Xiaobo Lu, Biao Lian, Gaurav Chaudhary, Benjamin~A. Piot, Giulio Romagnoli,
  Kenji Watanabe, Takashi Taniguchi, Martino Poggio, Allan~H. MacDonald,
  B.~Andrei Bernevig, and Dmitri~K. Efetov.
\newblock Multiple flat bands and topological hofstadter butterfly in twisted
  bilayer graphene close to the second magic angle.
\newblock {\em Proceedings of the National Academy of Sciences},
  118(30):e2100006118, July 2021.

\bibitem{Yaghi_2005}
Adrien~P. C{\^o}t{\'e}, Annabelle~I. Benin, Nathan~W. Ockwig, Michael
  O{\textquoteright}Keeffe, Adam~J. Matzger, and Omar~M. Yaghi.
\newblock Porous, crystalline, covalent organic frameworks.
\newblock {\em Science}, 310(5751):1166--1170, 2005.

\bibitem{Hahn_2019}
Seungyong Hahn, Kwanglok Kim, Kwangmin Kim, Xinbo Hu, Thomas Painter, Iain
  Dixon, Seokho Kim, Kabindra~R. Bhattarai, So~Noguchi, Jan Jaroszynski, and
  David~C. Larbalestier.
\newblock 45.5-tesla direct-current magnetic field generated with a
  high-temperature superconducting magnet.
\newblock {\em Nature}, 570(7762):496--499, June 2019.

\bibitem{CURATED_2019}
Daniele Ongari, Aliaksandr~V. Yakutovich, Leopold Talirz, and Berend Smit.
\newblock Building a consistent and reproducible database for adsorption
  evaluation in covalent{\textendash}organic frameworks.
\newblock {\em {ACS} Central Science}, 5(10):1663--1675, September 2019.

\bibitem{Dienstmaier_2011}
J\"{u}rgen~F. Dienstmaier, Alexander~M. Gigler, Andreas~J. Goetz, Paul Knochel,
  Thomas Bein, Andrey Lyapin, Stefan Reichlmaier, Wolfgang~M. Heckl, and Markus
  Lackinger.
\newblock Synthesis of well-ordered {COF} monolayers: Surface growth of
  nanocrystalline precursors versus direct on-surface polycondensation.
\newblock {\em {ACS} Nano}, 5(12):9737--9745, November 2011.

\bibitem{Dienstmaier_2012}
J\"{u}rgen~F. Dienstmaier, Dana~D. Medina, Mirjam Dogru, Paul Knochel, Thomas
  Bein, Wolfgang~M. Heckl, and Markus Lackinger.
\newblock Isoreticular two-dimensional covalent organic frameworks synthesized
  by on-surface condensation of diboronic acids.
\newblock {\em {ACS} Nano}, 6(8):7234--7242, July 2012.

\bibitem{Colson_2013}
John~W. Colson and William~R. Dichtel.
\newblock Rationally synthesized two-dimensional polymers.
\newblock {\em Nature Chemistry}, 5(6):453--465, May 2013.

\bibitem{Zhong_2019}
Yu~Zhong, Baorui Cheng, Chibeom Park, Ariana Ray, Sarah Brown, Fauzia Mujid,
  Jae-Ung Lee, Hua Zhou, Joonki Suh, Kan-Heng Lee, Andrew~J. Mannix, Kibum
  Kang, S.~J. Sibener, David~A. Muller, and Jiwoong Park.
\newblock Wafer-scale synthesis of monolayer two-dimensional porphyrin polymers
  for hybrid superlattices.
\newblock {\em Science}, 366(6471):1379--1384, November 2019.

\bibitem{Liu_2019}
Kejun Liu, Haoyuan Qi, Renhao Dong, Rishi Shivhare, Matthew Addicoat, Tao
  Zhang, Hafeesudeen Sahabudeen, Thomas Heine, Stefan Mannsfeld, Ute Kaiser,
  Zhikun Zheng, and Xinliang Feng.
\newblock On-water surface synthesis of crystalline, few-layer two-dimensional
  polymers assisted by surfactant monolayers.
\newblock {\em Nature Chemistry}, 11(11):994--1000, September 2019.

\bibitem{Li_2020}
Yusen Li, Weiben Chen, Guolong Xing, Donglin Jiang, and Long Chen.
\newblock New synthetic strategies toward covalent organic frameworks.
\newblock {\em Chemical Society Reviews}, 49(10):2852--2868, 2020.

\bibitem{Li2_2020}
Xing Li, Hai-Sen Xu, Kai Leng, See~Wee Chee, Xiaoxu Zhao, Noopur Jain, Hai Xu,
  Jingsi Qiao, Qiang Gao, In-Hyeok Park, Su~Ying Quek, Utkur Mirsaidov, and
  Kian~Ping Loh.
\newblock Partitioning the interlayer space of covalent organic frameworks by
  embedding pseudorotaxanes in their backbones.
\newblock {\em Nature Chemistry}, 12(12):1115--1122, November 2020.

\bibitem{OrtegaGuerrero_2021}
Andres Ortega-Guerrero, Hafeesudeen Sahabudeen, Alexander Croy, Arezoo Dianat,
  Renhao Dong, Xinliang Feng, and Gianaurelio Cuniberti.
\newblock Multiscale modeling strategy of 2d covalent organic frameworks
  confined at an air{\textendash}water interface.
\newblock {\em {ACS} Applied Materials {\&} Interfaces}, 13(22):26411--26420,
  May 2021.

\bibitem{Springer_2020}
Maximilian~A. Springer, Tsai-Jung Liu, Agnieszka Kuc, and Thomas Heine.
\newblock Topological two-dimensional polymers.
\newblock {\em Chemical Society Reviews}, 49(7):2007--2019, 2020.

\bibitem{Raptakis_2021}
Antonios Raptakis, Arezoo Dianat, Alexander Croy, and Gianaurelio Cuniberti.
\newblock Predicting the bulk modulus of single-layer covalent organic
  frameworks with square-lattice topology from molecular building-block
  properties.
\newblock {\em Nanoscale}, 13(2):1077--1085, 2021.

\bibitem{Dong_2016}
Liang Dong, Youngkuk Kim, Dequan Er, Andrew~M. Rappe, and Vivek~B. Shenoy.
\newblock Two-dimensional$\pi$-conjugated covalent-organic frameworks as
  quantum anomalous hall topological insulators.
\newblock {\em Physical Review Letters}, 116(9), February 2016.

\bibitem{Jiang_2021}
Wei Jiang, Xiaojuan Ni, and Feng Liu.
\newblock Exotic topological bands and quantum states in
  metal{\textendash}organic and covalent{\textendash}organic frameworks.
\newblock {\em Accounts of Chemical Research}, 54(2):416--426, January 2021.

\bibitem{Pham_2021}
Hung~Q. Pham and Nguyen-Nguyen Pham-Tran.
\newblock Topological insulating phase in single-layer pentagonal covalent
  organic frameworks: A reticular design using metal phthalocyanine.
\newblock {\em Chemistry of Materials}, June 2021.

\bibitem{Cui_2020}
Bin Cui, Xingwen Zheng, Jianfeng Wang, Desheng Liu, Shijie Xie, and Bing Huang.
\newblock Realization of lieb lattice in covalent-organic frameworks with
  tunable topology and magnetism.
\newblock {\em Nature Communications}, 11(1), January 2020.

\bibitem{Li_Lei_2020}
Jie Li, Lei Gu, and Ruqian Wu.
\newblock Transition-metal phthalocyanine monolayers as new chern insulators.
\newblock {\em Nanoscale}, 12(6):3888--3893, 2020.

\bibitem{Jiang_2020}
Wei Jiang, Shunhong Zhang, Zhengfei Wang, Feng Liu, and Tony Low.
\newblock Topological band engineering of lieb lattice in phthalocyanine-based
  metal{\textendash}organic frameworks.
\newblock {\em Nano Letters}, 20(3):1959--1966, February 2020.

\bibitem{Kuc_2020}
Agnieszka Kuc, Maximilian~A. Springer, Kamal Batra, Rosalba Juarez-Mosqueda,
  Christof W\"{o}ll, and Thomas Heine.
\newblock Proximity effect in crystalline framework materials: Stacking-induced
  functionality in {MOFs} and {COFs}.
\newblock {\em Advanced Functional Materials}, 30(41):1908004, February 2020.

\bibitem{Zhao_2020}
Chenfei Zhao, Hao Lyu, Zhe Ji, Chenhui Zhu, and Omar~M. Yaghi.
\newblock Ester-linked crystalline covalent organic frameworks.
\newblock {\em Journal of the American Chemical Society}, 142(34):14450--14454,
  August 2020.

\bibitem{Holec_2021}
Jan Holec, Beatrice Cogliati, James Lawrence, Alejandro Berdonces-Layunta,
  Pablo Herrero, Yuuya Nagata, Marzena Banasiewicz, Boleslaw Kozankiewicz,
  Martina Corso, Dimas~G. de~Oteyza, Andrej Jancarik, and Andre Gourdon.
\newblock A large starphene comprising pentacene branches.
\newblock {\em Angewandte Chemie International Edition}, 60(14):7752--7758,
  2021.

\bibitem{Pedersen_2013}
Jesper~Goor Pedersen and Thomas~Garm Pedersen.
\newblock Hofstadter butterflies and magnetically induced band-gap quenching in
  graphene antidot lattices.
\newblock {\em Physical Review B}, 87(23), June 2013.

\bibitem{Riao_2021}
Alberto Ria{\~{n}}o, Karol Struty{\'{n}}ski, Meng Liu, Craig~T. Stoppiello,
  Bel{\'{e}}n Lerma-Berlanga, Akinori Saeki, Carlos Mart{\'{\i}}-Gastaldo,
  Andrei~N. Khlobystov, Giovanni Valenti, Francesco Paolucci, Manuel
  Melle-Franco, and Aurelio Mateo-Alonso.
\newblock An expanded 2d fused aromatic network with 90-ring hexagons.
\newblock {\em Angewandte Chemie International Edition}, December 2021.

\bibitem{Slater_1954}
J.~C. Slater and G.~F. Koster.
\newblock Simplified lcao method for the periodic potential problem.
\newblock {\em Phys. Rev.}, 94:1498--1524, Jun 1954.

\bibitem{Manzano_2012}
H.~Manzano, A.~N. Enyashin, J.~S. Dolado, A.~Ayuela, J.~Frenzel, and
  G.~Seifert.
\newblock Do cement nanotubes exist?
\newblock {\em Advanced Materials}, 24(24):3239--3245, May 2012.

\bibitem{ASE_2017}
Ask~Hjorth Larsen, Jens~J{\o}rgen Mortensen, Jakob Blomqvist, Ivano~E Castelli,
  Rune Christensen, Marcin Du{\l}ak, Jesper Friis, Michael~N Groves, Bj{\o}rk
  Hammer, Cory Hargus, Eric~D Hermes, Paul~C Jennings, Peter~Bjerre Jensen,
  James Kermode, John~R Kitchin, Esben~Leonhard Kolsbjerg, Joseph Kubal,
  Kristen Kaasbjerg, Steen Lysgaard, J{\'{o}}n~Bergmann Maronsson, Tristan
  Maxson, Thomas Olsen, Lars Pastewka, Andrew Peterson, Carsten Rostgaard,
  Jakob Schi{\o}tz, Ole Sch\"{u}tt, Mikkel Strange, Kristian~S Thygesen, Tejs
  Vegge, Lasse Vilhelmsen, Michael Walter, Zhenhua Zeng, and Karsten~W
  Jacobsen.
\newblock The atomic simulation environment{\textemdash}a python library for
  working with atoms.
\newblock {\em Journal of Physics: Condensed Matter}, 29(27):273002, June 2017.

\bibitem{pybinding}
Dean Moldovan, Mi\v{s}a Anđelkovi\'{c}, and Francois Peeters.
\newblock {pybinding v0.9.5: a Python package for tight- binding calculations},
  August 2020.
\newblock {This work was supported by the Flemish Science Foundation (FWO-Vl)
  and the Methusalem Funding of the Flemish Government.}

\bibitem{Peierls_1933}
R.~Peierls.
\newblock Zur theorie des diamagnetismus von leitungselektronen.
\newblock {\em Zeitschrift für Physik}, 80(11-12):763--791, November 1933.

\bibitem{KPM_method}
Alexander Wei\ss{}e, Gerhard Wellein, Andreas Alvermann, and Holger Fehske.
\newblock The kernel polynomial method.
\newblock {\em Rev. Mod. Phys.}, 78:275--306, Mar 2006.

\bibitem{DFTB_2020}
B.~Hourahine, B.~Aradi, V.~Blum, F.~Bonaf{\'{e}}, A.~Buccheri, C.~Camacho,
  C.~Cevallos, M.~Y. Deshaye, T.~Dumitric{\u{a}}, A.~Dominguez, S.~Ehlert,
  M.~Elstner, T.~van~der Heide, J.~Hermann, S.~Irle, J.~J. Kranz,
  C.~K\"{o}hler, T.~Kowalczyk, T.~Kuba{\v{r}}, I.~S. Lee, V.~Lutsker, R.~J.
  Maurer, S.~K. Min, I.~Mitchell, C.~Negre, T.~A. Niehaus, A.~M.~N. Niklasson,
  A.~J. Page, A.~Pecchia, G.~Penazzi, M.~P. Persson,
  J.~{\v{R}}ez{\'{a}}{\v{c}}, C.~G. S{\'{a}}nchez, M.~Sternberg, M.~St\"{o}hr,
  F.~Stuckenberg, A.~Tkatchenko, V.~W. z.~Yu, and T.~Frauenheim.
\newblock {DFTB+}, a software package for efficient approximate density
  functional theory based atomistic simulations.
\newblock {\em The Journal of Chemical Physics}, 152(12):124101, March 2020.

\end{thebibliography}

\noindent \textsf{\textbf{Acknowledgements}}\\
The authors acknowledge financial support from the DFG project CRC-1415 (No. 417590517). The authors thank Antonios Raptakis for providing the geometry-files of the COF structures.\\\\
\noindent \textsf{\textbf{Author contribution}}\\
D.B. performed the calculations. R.B. and G.C. supervised and guided the project. All authors contributed equally to the preparation of this manuscript.\\\\
\noindent \textsf{\textbf{Competing interests}}\\
The authors declare no competing interests.\\

\end{document}